\documentclass[a4paper,12pt]{article}

\usepackage{amssymb}
\usepackage[dvips]{color}
\usepackage[latin1]{inputenc}
\usepackage{graphicx}
\usepackage{gensymb}
\usepackage{amsmath}
\newcommand{\bsubeqs}{\begin{subequations}}
\newcommand{\esubeqs}{\end{subequations}}

\title{An effective-gravity perspective on the Sun-Jupiter-comet three-body system}

\date{\today}

\author{Emmanuele Battista ORCID: 0000-0001-5361-7109 \\
Institute for Theoretical Physics, \\
Karlsruhe Institute
of Technology (KIT), \\
76128 Karlsruhe, Germany \\
Institute for Nuclear Physics, \\
Karlsruhe Institute of Technology (KIT), \\
Hermann-von-Helmholtz-Platz 1, \\
76344 Eggenstein-Leopoldshafen, Germany \and
Giampiero Esposito ORCID: 0000-0001-5930-8366 \\
Istituto Nazionale di Fisica Nucleare, Sezione di Napoli, \\
Complesso Universitario di Monte S. Angelo, \\
Via Cintia Edificio 6, 80126 Napoli, Italy \and
Angelo Tartaglia ORCID: 0000-0001-9763-1615 \\
Istituto Nazionale di Astrofisica, \\
Osservatorio Astronomico di Torino, \\
Via Osservatorio 20, 10025 Pino Torinese (TO), Italy}

\begin{document}
\maketitle

\begin{abstract}
Within the solar system, approximate realizations of the three-body problem occur when a comet
approaches a planet while being affected mainly by such a planet and the Sun,
and this configuration was investigated by Tisserand within
the framework of Newtonian gravity. The exact relativistic
treatment of the problem is not an easy task, but
the present paper develops an approximate calculational
scheme which computes for the first time the tiny effective-gravity correction
to the equation of the surface for all points of which it is equally advantageous to regard
the heliocentric motion as being perturbed by the attraction of Jupiter, or the jovicentric
motion as being perturbed by the attraction of the Sun.
This analysis completes the previous theoretical investigations of
effective-gravity corrections to the Newtonian analysis of
three-body systems, and represents an intermediate step towards
relativistic effects on cometary motions.
\end{abstract}

\section{Introduction}
\setcounter{equation}{0}

A complete understanding of potentialities, applications and limits of Newton's and Einstein's
theories of gravity has required dedicated efforts along more than three centuries, by now. For
example, at the end of nineteenth century the monumental treatise by Tisserand on celestial
mechanics \cite{Tisserand} presented in great detail the work of d'Alembert and Laplace on the
motion of comets when they are approaching a planet. This analysis stimulated Fermi himself,
when he wrote his Scuola Normale Superiore dissertation \cite{Fermi}, devoted to an investigation
of cometary orbits with the help of probability theory and of the classical theory of
restricted three-body problems.

What has motivated our research has been therefore, on the one hand, the many (recent) investigations
of three-body problems in general relativity \cite{Krefetz,Bhat,YA1,YA2,YA3,YA4}
and effective-field-theory models of gravity \cite{BE1,BE2,BE3,BE4,BA,D1,D2,D3,D4,D5,D6}
(see also \cite{Alshaery2020} for a new original approach), and on
the other hand the consideration that the passage of comets provides in the solar system some
very interesting realizations of three-body systems in celestial mechanics.
Short-period comets \cite{Duncan} are
thought to generate in the Kuiper belt and have predictable orbits with short periods, i.e.,
up to $200$ years. Two major families of short-period comets are the {\it Jupiter family} with
periods of less than $20$ years and the Halley family with periods in between $20$ and $200$ years.
Interestingly, even though their orbits can be predicted with some accuracy, some of these
short-period comets might be gravitationally perturbed and become long-period objects. More
precisely, gravitational effects of the outer planets can cause these bodies to alter their paths
into highly elliptical orbits that take them close to the Sun. Long-period comets are instead
thought to generate in the Oort cloud and have unpredictable orbits, with periods much longer
than $200$ years. Their detection is extremely difficult for mankind because they can return
on their steps after thousands or even millions of year (or not at all).

Several processes deserve careful consideration, e.g.,
the capture of comets with parabolic orbit by Jupiter \cite{Havnes}.
Another intriguing difficulty is the choice of the appropriate formalism for our analysis.
In the last decade of the twentieth century, the outstanding work by Damour, Soffel and Xu
\cite{DSX1,DSX2,DSX3,DSX4,DSX5} led indeed to a complete prescription for studying equations
of motion of $N$ bodies in celestial mechanics, which involve a repeated application of the
inverse of the wave operator and are therefore nonlocal. Since, for a three-body problem, we
are still far from the level of accuracy reached for relativistic binary systems
\cite{BD1,BD2,DB1}, we here resort first to a shortcut,
i.e., an approximation method, a hybrid scheme,
which is nevertheless of physical interest, to the same extent that ordinary quantum
mechanics, despite not being a relativistic quantum theory, is of much help in
evaluating bound states and transition probabilities. The milestones we rely upon are
as follows.
\vskip 0.3cm
\noindent
(i) In the effective-field-theory approach to quantum gravity, one discovers that the
Newtonian potential between two bodies of masses $M_{A}$ and $M_{B}$ should be replaced
by the expression \cite{D1,D3,D5,BE4}
\begin{equation}
V_{E}(r)=-{G M_{A} M_{B} \over r_{AB}}
\left[1+\kappa_{1}{(L_{A}+L_{B})\over r_{AB}}
+\kappa_{2}\left({L_{P}\over r_{AB}}\right)^{2}
+{\rm O}(G^{2})\right],
\label{(1.1)}
\end{equation}
where $L_{A}$ and $L_{B}$ are their gravitational radii
\begin{equation}
L_{A} \equiv {G M_{A}\over c^{2}}, \;
L_{B} \equiv {G M_{B}\over c^{2}},
\label{(1.2)}
\end{equation}
$L_{P}$ is the Planck length
\begin{equation}
L_{P} \equiv \sqrt{G \hbar \over c^{3}},
\label{(1.3)}
\end{equation}
the length $r_{AB}$ is the mutual distance between the bodies, while $\kappa_{1}$ and $\kappa_{2}$
are dimensionless parameters obtained either from a detailed application of Feynman diagrams' technique
\cite{D1,D2,D3} or the modern on-shell unitary-based methods \cite{Bern1994, Bohr2014}.
It should be stressed that, although the numerical effect of $L_{P}$ is
immaterial in the calculations devoted to Lagrangian points and their stability \cite{BE1,BE2,BE3,BE4},
the dependence of $\kappa_{1}$ on $\kappa_{2}$ implies that
{\it the weight factor $\kappa_{1}$ for the classical
gravitational radii depends actually on the underlying quantum world}.
\vskip 0.3cm
\noindent
(ii) The formula (\ref{(1.1)}) tells us that quantum corrections map the dimensionless ratio
\begin{equation}
U_{A} \equiv {L_{A}\over r_{AB}},
\label{(1.4)}
\end{equation}
into \cite{BE4}
\begin{equation}
V_{A} \sim \left[1+\kappa_{2} \left({L_{P}\over r_{AB}}\right)^{2}\right]U_{A}
+\kappa_{1}(U_{A})^{2}.
\label{(1.5)}
\end{equation}
Thus, if (\ref{(1.5)}) is applied to replace $U_{A}$ and
$U_{B}$ in the Newtonian formula of effective
potential $W_{\rm eff}$ for circular restricted three-body problem, one finds
in $c=1$ units \cite{BE4}, denoting by $\omega$ the angular frequency for rotation of the
$\xi,\eta$ axes about the $\zeta$ axis \cite{Bhat},
\begin{eqnarray}
W_{\rm eff}&=&{\omega^{2}\over 2} (\xi^{2}+\eta^{2})+\left[U_{A}+U_{B}
+\kappa_{1}\left((U_{A})^{2}+(U_{B})^{2}\right)\right]
\nonumber \\
&+& {\rm O}(G^{2}).
\label{(1.6)}
\end{eqnarray}
Remarkably, the general relativity formula in $c=1$ units is instead \cite{Bhat,BE4}
\begin{eqnarray}
W_{\rm eff} & \sim & {\omega^{2}\over 2}(\xi^{2}+\eta^{2})
+\left[U_{A}+U_{B}-{1 \over 2} \left((U_{A})^{2}+(U_{B})^{2}\right)\right]
\nonumber \\
&+& \; {\rm remainder},
\label{(1.7)}
\end{eqnarray}
and the first lines of (\ref{(1.6)}) and (\ref{(1.7)}) agree if $\kappa_{1}=-{1 \over 2}$.
The value $\kappa_{1}=-{1 \over 2}$ is indeed allowed by the effective-field-theory
approach to quantum gravity, and it corresponds to the so-called bound-state option
for the underlying Feynman diagrammatics \cite{D1,D3,D5,BE4}. This means that, upon
application of (\ref{(1.1)}) or (\ref{(1.5)}) to the
Newtonian formulas for three-body problems, one
can obtain many of the relevant terms of the associated effective potential, so that
a valuable recipe is available. According to it,
one can insert (\ref{(1.1)}) (or (\ref{(1.5)}))
in all Newtonian formulas, and obtain valuable information on the otherwise (too) lengthy
fully relativistic calculations. Bearing in mind also this feature, which was first pointed
out in Ref. \cite{BE4}, the plan of our paper is as follows.

Section \ref{Sec_Cometary_Newton} describes the Newtonian
formulation of perturbations of cometary motions
when comets come very close to planets. Section
\ref{Sec_perturb_EFT} finds the leading relativistic
corrections of this treatment with the help of the map (\ref{(1.1)}), by focusing on the equation
of the surface for all points of which it is equally advantageous to view the heliocentric
motion as being perturbed by the attraction of Jupiter, or the jovicentric motion as
being perturbed by the attraction of the Sun.
Concluding remarks are made in Sec. \ref{Sec_conclusions}.

\section{Perturbations of cometary motions in Newtonian gravity} \label{Sec_Cometary_Newton}
\setcounter{equation}{0}

Following the monograph of Tisserand \cite{Tisserand}, we denote by
$(x,y,z)$ the heliocentric rectangular coordinates of the comet, by $(x',y',z')$ the coordinates
of the planet, here taken to be Jupiter, and by $(\xi,\eta,\zeta)$ the jovicentric coordinates
of the comet, with respect to axes that are parallel to the fixed axes. Moreover,
$m_{\odot}=2 \times 10^{30}$ Kg is the mass of the Sun, $m'$ is the mass of Jupiter:
\begin{equation}
{m' \over m_{\odot}}={1 \over 1047},
\label{(2.1)}
\end{equation}
$\rho$ is the Euclidean distance comet-Jupiter, $r,r'$ the Euclidean distances
comet-Sun and Jupiter-Sun, respectively. The Newtonian equations of motion
turn out to be \cite{Tisserand}
\begin{equation}
\left[{d^{2} \over dt^{2}}+{G m_{\odot} \over r^{3}}\right]x
=Gm' \left({(x'-x)\over \rho^{3}}-{x' \over {r'}^{3}}\right),
\label{(2.2)}
\end{equation}
\begin{equation}
\left[{d^{2} \over dt^{2}}+{G m_{\odot} \over r^{3}}\right]y
=Gm' \left({(y'-y)\over \rho^{3}}-{y' \over {r'}^{3}}\right),
\label{(2.3)}
\end{equation}
\begin{equation}
\left[{d^{2} \over dt^{2}}+{G m_{\odot} \over r^{3}}\right]z
=Gm' \left({(z'-z)\over \rho^{3}}-{z' \over {r'}^{3}}\right),
\label{(2.4)}
\end{equation}
\begin{equation}
\left({d^{2} \over dt^{2}}+\Omega^2\right)x'=0,
\label{(2.5)}
\end{equation}
\begin{equation}
\left({d^{2} \over dt^{2}}+\Omega^2\right)y'=0,
\label{(2.6)}
\end{equation}
\begin{equation}
\left({d^{2} \over dt^{2}}+\Omega^2\right)z'=0,
\label{(2.7)}
\end{equation}
where
\begin{equation}
\Omega^2 = {G(m_{\odot}+m') \over {r'}^{3}}.
\label{newtonian_Omega}
\end{equation}
The unprimed, primed and Greek lower case coordinates are
related by linear equations, i.e.
\begin{equation}
x=x'+\xi, \; y=y'+\eta, \; z=z'+\zeta,
\label{(2.8)}
\end{equation}
and by virtue of (\ref{(2.2)})-(\ref{(2.8)}) one obtains the
system of second-order equations
\begin{equation}
\left[{d^{2} \over dt^{2}}+{G m' \over \rho^{3}}\right] \xi
=G m_{\odot}\left({x' \over {r'}^{3}}-{x \over r^{3}}\right),
\label{(2.9)}
\end{equation}
\begin{equation}
\left[{d^{2} \over dt^{2}}+{G m' \over \rho^{3}}\right] \eta
=G m_{\odot}\left({y' \over {r'}^{3}}-{y \over r^{3}}\right),
\label{(2.10)}
\end{equation}
\begin{equation}
\left[{d^{2} \over dt^{2}}+{G m' \over \rho^{3}}\right] \zeta
=G m_{\odot}\left({z' \over {r'}^{3}}-{z \over r^{3}}\right).
\label{(2.11)}
\end{equation}
Equations (\ref{(2.2)})-(\ref{(2.4)}) pertain to the
heliocentric motion of the comet, whereas
(\ref{(2.5)})-(\ref{newtonian_Omega}) refer to the
elliptical motion of Jupiter about the barycentre
of the system, which for simplicity coincides exactly
with the position of the Sun; let now $R$ be
the modulus of the force per unit mass resulting from the attraction of the Sun, and let
$F$ be the modulus of the perturbing force per unit mass. They are given by \cite{Tisserand}
\begin{equation}
R={G m_{\odot}\over r^{2}},
\label{(2.12)}
\end{equation}
\begin{equation}
F=G m' \sqrt{\left({(x'-x)\over \rho^{3}}-{x' \over {r'}^{3}}\right)^{2}
+\left({(y'-y)\over \rho^{3}}-{y'\over {r'}^{3}}\right)^{2}
+\left({(z'-z)\over \rho^{3}}-{z' \over {r'}^{3}}\right)^{2}}.
\label{(2.13)}
\end{equation}
Equations (\ref{(2.9)})-(\ref{(2.11)}) pertain instead to the jovicentric motion produced by the
attraction $R'$ of Jupiter, and the perturbing force exerted by the Sun. The modulus
of such forces per unit mass reads as
\begin{equation}
R'={G m' \over \rho^{2}},
\label{(2.14)}
\end{equation}
and
\begin{equation}
F'=G m_{\odot} \sqrt{
\left({x' \over {r'}^{3}}-{x \over r^{3}}\right)^{2}
+\left({y' \over {r'}^{3}}-{y \over r^{3}}\right)^{2}
+\left({z' \over {r'}^{3}}-{z \over r^{3}}\right)^{2}},
\label{(2.15)}
\end{equation}
respectively.

\subsection{Heliocentric vs. jovicentric motion}

A concept of crucial importance is expressed by the condition \cite{Tisserand}
\begin{equation}
{F \over R}={F' \over R'},
\label{(2.16)}
\end{equation}
which defines implicitly a surface for all points of which it is equally advantageous
to regard the heliocentric motion as being perturbed by the attraction of Jupiter, or
the jovicentric motion as being perturbed by the attraction of the Sun. On denoting
simply by $m'$ the ratio in (\ref{(2.1)}),
condition (\ref{(2.16)}) reads eventually as \cite{Tisserand}
\begin{equation}
{m'}^{2}r^{2}
\sqrt{{1 \over \rho^{4}}+{1 \over {r'}^{4}}
+2{(x' \xi + y' \eta + z' \zeta) \over \rho^{3} {r'}^{3}}}
=\rho^{2} \sqrt{{1 \over r^{4}}+{1 \over {r'}^{4}}
-2{(xx'+yy'+zz')\over r^{3}{r'}^{3}}},
\label{(2.17)}
\end{equation}
where we have exploited the identities
\begin{equation}
r^{2}=x^{2}+y^{2}+z^{2}, \;
{r'}^{2}={x'}^{2}+{y'}^{2}+{z'}^{2}, \;
\rho^{2}=\xi^{2}+\eta^{2}+\zeta^{2}.
\label{(2.18)}
\end{equation}
At this stage, it is convenient to introduce the variables $\theta$ and $u$ by means
of the definitions \cite{Tisserand}
\begin{equation}
\cos(\theta) \equiv {(x' \xi+y' \eta + z' \zeta)\over \rho r'},
\label{(2.19)}
\end{equation}
\begin{equation}
u \equiv {\rho \over r'}.
\label{(2.20)}
\end{equation}
The dimensionless parameter $u$ approaches $0$, since it is only at short distance from
Jupiter that the transformation considered can be convenient. Moreover, by virtue of the
linear relations (\ref{(2.8)}), a term on the right-hand side of (\ref{(2.17)}) reads as
\begin{equation}
xx'+yy'+zz'={r'}^{2}(1+u \cos (\theta)),
\label{(2.21)}
\end{equation}
while the squared distance $r^{2}$ can be expressed in the form
\begin{equation}
r^{2}=(x'+\xi)^{2}+(y'+\eta)^{2}+(x'+\zeta)^{2}
={r'}^{2}(1+2 u \cos(\theta)+u^{2}).
\label{(2.22)}
\end{equation}
In light of (\ref{(2.19)})-(\ref{(2.22)}), we can express the squared
Jupiter mass in (\ref{(2.17)}) by means of
$u$ and $\theta$ only, after writing $\rho=u \; r'$ and then expressing ${r' \over r}$
from (\ref{(2.22)}). Hence one finds \cite{Tisserand}
\begin{equation}
{m'}^{2}={u^{4}\over (1+2u \cos (\theta)+u^{2})^{2}
\sqrt{1+2u^{2}\cos(\theta)+u^{4}}} \sqrt{P(u,\theta)}.
\label{(2.23)}
\end{equation}
The exact form of the function $P$ is \cite{Tisserand}
\begin{equation}
P(u,\theta) \equiv 1+(1+2u \cos(\theta)+u^{2})^{2}
-2 \sqrt{1+2u \cos(\theta)+u^{2}} \; (1+u \cos(\theta)).
\label{(2.24)}
\end{equation}
Since $u$ approaches $0$ in our physical model, we only need the small-$u$
expansion of $P(u,\theta)$. For this purpose, we first notice that, at fixed $\theta$,
\begin{eqnarray}
\; & \; & f(u) \equiv \sqrt{1+2u \cos(\theta)+u^{2}}
\nonumber \\
&=& 1+u \cos(\theta)+{1 \over 2}\sin^{2}(\theta)u^{2}
-{1 \over 2}\cos(\theta) \sin^{2}(\theta)u^{3}+{\rm O}(u^{4}),
\label{(2.25)}
\end{eqnarray}
and hence we find, after following patiently a number of exact or partial cancellations,
\begin{equation}
P(u,\theta)=u^{2}[1+3 \cos^{2} (\theta)]+4u^{3}\cos(\theta)+{\rm O}(u^{4}).
\label{(2.26)}
\end{equation}
By virtue of (\ref{(2.23)}) and (\ref{(2.26)}), we obtain first the
approximate formula \cite{Tisserand}
\begin{eqnarray}
{m'}^{2} &=& {u^{4} \sqrt{u^{2}[1+3 \cos^{2} (\theta)]
+4 u^{3}\cos (\theta)+{\rm O}(u^{4})} \over
(1+4u \cos(\theta)+{\rm O}(u^{2}))(1+{\rm O}(u^{2}))}
\nonumber \\
&=& u^{5}(1-4u \cos(\theta)+{\rm O}(u^{2}))
\sqrt{1+3 \cos^{2} (\theta)+4u \cos (\theta)}
\nonumber \\
&=& u^{5} \sqrt{1+3 \cos^{2} (\theta)}
\left(1-2u \cos(\theta){[1+6 \cos^{2} (\theta)]\over
[1+3 \cos^{2} (\theta)]}
+{\rm O}(u^{2}) \right),
\label{(2.27)}
\end{eqnarray}
and eventually we solve approximately for the dimensionless parameter $u$
in the form \cite{Tisserand}
\begin{eqnarray}
u &=& \left({{m'}^{2}\over \sqrt{1+3 \cos^{2} (\theta)}}\right)^{1 \over 5}
\left(1+2 u \cos (\theta)
{[1+6 \cos^{2} (\theta)]\over
[1+3 \cos^{2} (\theta)]}\right)^{1 \over 5}
\nonumber \\
& \approx &
\left({{m'}^{2}\over \sqrt{1+3 \cos^{2} (\theta)}}\right)^{1 \over 5}
\nonumber \\
&+& {2 \over 5}\cos (\theta)
\left({{m'}^{2}\over \sqrt{1+3 \cos^{2} (\theta)}}\right)^{2 \over 5}
{[1+6 \cos^{2} (\theta)]\over
[1+3 \cos^{2} (\theta)]}.
\label{(2.28)}
\end{eqnarray}
By virtue of (\ref{(2.1)}), one has in all cases
\begin{equation}
\left({{m'}^{2}\over \sqrt{1+3 \cos^{2} (\theta)}}\right)^{1 \over 5}
< {m'}^{2 \over 5}=0.062,
\label{(2.29)}
\end{equation}
and hence one can safely use the approximate formula
\begin{equation}
\rho=r' \left({{m'}^{2}\over \sqrt{1+3 \cos^{2} (\theta)}}\right)^{1 \over 5}.
\label{(2.30)}
\end{equation}
This is the approximate equation of the desired surface in polar coordinates $\rho$
and $\theta$, the polar axis being obtained by extending the line $SP$ joining the Sun
with the planet, having set the origin at the centre of the planet. This is a
surface of revolution about the $SP$ line, and does not differ much from a sphere,
since $\rho$ varies in between the two limits
$$
{m'}^{2 \over 5} \; r' \; \; {\rm and} \; \;
{{m'}^{2 \over 5}\over 2^{1 \over 5}} \; r',
$$
which correspond to $\theta={\pi \over 2}$ and $\theta=0$, respectively, and whose
ratio equals $1.15$. One can therefore say, with little error, that the surface defined by
the condition (\ref{(2.16)}) is a sphere of radius ${m'}^{2 \over 5} \; r'$, called by
Laplace the {\it sphere of influence} of the planet.\footnote{Strictly, Laplace considered the
fifth root of half the square of $m'$, as pointed out by Tisserand \cite{Tisserand}.}
Outside of such a sphere, one has ${F \over R}< {F' \over R'}$, and it is therefore
advantageous to start from the heliocentric motion of the comet, and to evaluate
the perturbations caused by the planet. Within the sphere of influence, one has instead
${F' \over R'}<{F \over R}$, and it is more advantageous to consider the jovicentric motion,
and to evaluate as a next step the perturbations resulting from the Sun.

\section{Perturbations of cometary motions in effective field
theories of gravity} \label{Sec_perturb_EFT}
\setcounter{equation}{0}

The quantum effects considered in (\ref{(1.1)}) affect the potential,
whereas the Newtonian model outlined in Sec. \ref{Sec_Cometary_Newton}
relies upon the evaluation of forces. Thus, we need to
propose first a modified force formula in order to
write down the effective-gravity counterpart of Sec. \ref{Sec_Cometary_Newton}.
For this purpose, we assume that we can still
express the force as minus the gradient of the effective potential, i.e.,
\begin{equation}
F_{k}=(-{\rm grad} \quad V_{E})_{k}, \quad k=1,2,3, \quad x_{1}=x,x_{2}=y,x_{3}=z,
\label{(3.1)}
\end{equation}
which implies that (since ${\partial r \over \partial x_{k}}={x_{k}\over r}$)
\begin{equation}
F_{k}=-{G M_{A}M_{B}\over r^{3}}x_{k}\left[1
+2 \kappa_{1}{(L_{A}+L_{B})\over r}
+3 \kappa_{2} \left({L_{P}\over r}\right)^{2}
+{\rm O}(G^{2})\right].
\label{(3.2)}
\end{equation}
The Newtonian formulas (\ref{(2.12)}) and (\ref{(2.13)}) for the modulus of
perturbing forces receive therefore an additional contribution from $\kappa_{1}$
according to the prescriptions (we use the subscript $E$ to denote the influence of
effective-gravity calculations, and we neglect the gravitational radius $L_{J}$ of Jupiter
with respect to the gravitational radius $L_{S}$ of the Sun)
\begin{equation}
R_{E} \sim {G m_{\odot}\over r^{2}}\left[1
+2 \kappa_{1}{L_{S}\over r}
+{\rm O}(L_{P}^{2})\right],
\label{(3.3)}
\end{equation}
\begin{equation}
F_{E} \sim Gm' \sqrt{\sum_{k=1}^{3}
\left[
{(x_{k}'-x_{k})\over \rho^{3}}\left(1+2 \kappa_{1}{L_{J}\over \rho}\right)
-{x_{k}'\over {r'}^{3}}\left(1+2 \kappa_{1}{L_{S}\over r'}\right)+{\rm O}(L_{P}^{2})\right]^{2}},
\label{(3.4)}
\end{equation}
where $x_{1}'=x',x_{2}'=y',x_{3}'=z'$. Now we point out that, up to ${\rm O}(L_{P}^{2})$ terms,
\begin{eqnarray}
\; & \; &
\left[
{(x_{k}'-x_{k})\over \rho^{3}}\left(1+2 \kappa_{1}{L_{J}\over \rho}\right)
-{x_{k}'\over {r'}^{3}}\left(1+2 \kappa_{1}{L_{S}\over r'}\right)\right]^{2}
\nonumber \\
&=& \left({(x_{k}'-x_{k})\over \rho^{3}}-{x_{k}' \over {r'}^{3}}\right)^{2}
+4 \kappa_{1}\left({(x_{k}'-x_{k})\over \rho^{3}}-{x_{k}'\over {r'}^{3}}\right)
\left({L_{J}\over \rho}{(x_{k}'-x_{k})\over \rho^{3}}
-{L_{S}\over r'}{x_{k}'\over {r'}^{3}}\right)
\nonumber \\
&+& {\rm O}(G^{2}),
\label{(3.5)}
\end{eqnarray}
and hence we find
\begin{equation}
F_{E} \sim Gm' \sqrt{P_{1}},
\label{(3.6)}
\end{equation}
where
\begin{eqnarray}
P_{1}(u,\theta)&=& {1 \over \rho^{4}}+{1 \over {r'}^{4}}
+2{(x' \xi + y' \eta + z' \zeta)\over \rho^{3}{r'}^{3}}
\nonumber \\
&+& 4 \kappa_{1} \sum_{k=1}^{3}\left[
\left({(x_{k}'-x_{k})\over \rho^{3}}-{x_{k}'\over {r'}^{3}}\right)
\left({L_{J}\over \rho}{(x_{k}'-x_{k})\over \rho^{3}}
-{L_{S}\over {r'}} {{x_{k}'}\over {r'}^{3}}\right)\right]+{\rm O}(G^{2})
\nonumber \\
&=& {1 \over \rho^{4}}+{1 \over {r'}^{4}}
+{2 \over (\rho r')^{2}}
\cos (\theta) +4 \kappa_{1} f_{1} + {\rm O}(G^{2}),
\label{(3.7)}
\end{eqnarray}
having set (see (\ref{(2.8)}))
\begin{equation}
f_{1} \equiv  \sum_{k=1}^{3} \left[
\left({\xi_k \over \rho^{3}}
+{x'_k \over {r'}^{3}}\right)
\left({L_{J}\over \rho}{\xi_k \over \rho^{3}}
+{L_{S}\over r'}{x'_k \over {r'}^{3}}\right)\right],
\label{(3.8)}
\end{equation}
with $\xi_1=\xi$, $\xi_2=\eta$, $\xi_3=\zeta$.

With analogous procedure, we propose to replace the Newtonian formulas
(\ref{(2.14)}) and (\ref{(2.15)}) with their effective-gravity counterparts
\begin{equation}
R_{E}' \sim {G m' \over \rho^{2}} \left(1+2 \kappa_{1}
{L_{J} \over \rho}+{\rm O}(L_{P}^{2})\right),
\label{(3.9)}
\end{equation}
\begin{eqnarray}
F_{E}' & \sim & G m_{\odot} \sqrt{\sum_{k=1}^{3}\left[
{x_{k}' \over {r'}^{3}}-{x_{k}\over r^{3}}
+2 \kappa_{1} \left({L_{S} \over r'}{x_{k}'\over {r'}^{3}}
-{L_{S}\over r}{x_{k}\over r^{3}}\right)+{\rm O}(L_{P}^{2})\right]^{2}}.
\label{(3.10)}
\end{eqnarray}
Therefore, we can write
\begin{equation}
F_{E}' \sim G m_{\odot}\sqrt{P_{2}},
\end{equation}
where up to ${\rm O}(L_{P}^{2})$ (see (\ref{(2.21)}))
\begin{equation}
P_{2}(u,\theta)={1 \over {r'}^{4}}+{1 \over r^{4}}
-2{{r'}^{2}(1+u \cos (\theta))\over r^{3}{r'}^{3}}
+4 \kappa_{1}L_{S}f_{2} + {\rm O}(G^{2}),
\label{(3.11)}
\end{equation}
having set
\begin{equation}
f_{2} \equiv \sum_{k=1}^3 \left[
\left({x'_k \over {r'}^{3}}-{x_k \over r^{3}}\right)
\left({x'_k \over {r'}^{4}}-{x_k \over r^{4}}\right)\right].
\label{(3.12)}
\end{equation}

The effective-gravity counterpart of condition (\ref{(2.16)}), i.e.,
\begin{equation}
{F_{E}\over R_{E}}={{F_{E}'} \over {R_{E}'}},
\label{(3.13)}
\end{equation}
leads therefore to the equation (cf. (\ref{(2.17)}) and (\ref{(2.23)}))
\begin{equation}
{m'}^{2}= \left({\rho\over r}\right)^{2}
\left(1+2 \kappa_{1}{L_{S}\over r}+{\rm O}(L_{P}^{2})\right)
\left(1+2 \kappa_{1}{L_{J}\over \rho}+{\rm O}(L_{P}^{2})\right)^{-1}
\sqrt{P_2(u,\theta)\over P_1(u,\theta)}.
\label{(3.14)}
\end{equation}
At this stage, by exploiting Eqs. (\ref{(2.8)}) and (\ref{(2.18)})-(\ref{(2.22)})
we can rewrite Eqs. (\ref{(3.8)}) and (\ref{(3.12)}) as follows
(the calculations below are both new and very instructive, hence they
deserve our special care):
\begin{eqnarray}
f_{1}&=& {1 \over {r'}^{7}} \sum_{k=1}^3\left[
\left(x'_k+{\xi_k \over u^{3}}\right)
\left(L_{S}x'_k+{L_{J}\xi_k \over u^{4}}\right)\right]
\nonumber \\
&=& {1 \over \rho^{7}}
\Bigr[L_{J}\rho^{2}+(uL_{S}+L_{J})\rho r' \cos(\theta) u^{3}
+L_{S}{r'}^{2}u^{7}\Bigr]
\nonumber \\
&=& {1 \over \rho^{5}}\Bigr[L_{J}+(uL_{S}+L_{J})\cos (\theta) u^{2}
+L_{S}u^{5}\Bigr],
\label{(3.17)}
\end{eqnarray}
\begin{eqnarray}
f_{2} &=& {1 \over {r'}^{7}}\sum_{k=1}^3\left\{
\left[x'_k-\gamma^{-{3 \over 2}}(x'_k+\xi_k)\right]
\left[x_k'-\gamma^{-2}(x'_k+\xi_k)\right]\right\}
\nonumber \\
&=& \dfrac{1}{r^{\prime 7}}\left\{ r^{\prime 2}
+ \gamma^{-{7 \over 2}} \left[ r^2 -\left(r^{\prime 2}
+\rho r^{\prime}\cos(\theta) \right)\left(\gamma^{{3 \over 2}}
+\gamma^2\right) \right]\right \}
\nonumber \\
&=& {u^{5}\over \rho^{5}} \left \{
1+ \gamma^{-{5 \over 2}} \Bigr[1-\left(\gamma^{1 \over 2}+\gamma\right)
\left(1+u \cos \left(\theta\right)\right) \Bigr] \right \},
\label{(3.18)}
\end{eqnarray}
where
\begin{equation}
\gamma \equiv 1 +2u \cos(\theta)+u^{2}.
\label{(3.19)}
\end{equation}
In view of the forthcoming calculations, it will be useful to write (\ref{(3.18)}) as
\begin{equation}
f_2=\left(\dfrac{u}{\rho}\right)^5 B(\gamma),
\label{def_f2}
\end{equation}
with
\begin{equation}
B(\gamma) \equiv 1+ \gamma^{-{5 \over 2}}
\Bigr[1-\left(\gamma^{1 \over 2}+\gamma\right)
\left(1+u \cos \left(\theta\right)\right) \Bigr],
\end{equation}
and to take into account that
\begin{equation}
{\rho \over r}=u {r' \over r}
=u \; \gamma^{-{1 \over 2}}.
\label{(3.20)}
\end{equation}
Bearing in mind the above calculations, we can write Eq. (\ref{(3.14)}) as
\begin{equation}
{m'}^{2}= \dfrac{u^2}{\gamma}
\dfrac{\left(1+2 \kappa_{1}{L_{S}\over r}+{\rm O}(L_{P}^{2})\right)}
{\left(1+2 \kappa_{1}{L_{J}\over \rho}
+{\rm O}(L_{P}^{2})\right)} \sqrt{P_2(u,\theta)\over P_1(u,\theta)}.
\end{equation}
Inspired by the classical calculations of Sec.
\ref{Sec_Cometary_Newton}, we extract the factor $1/r^4$
and $1/\rho^4$ from $P_2(u,\theta)$ and $P_1(u,\theta)$,
respectively, and hence on exploiting
(\ref{(3.20)}) the previous equation can be expressed in the form
\begin{equation}
{m'}^{2}= \dfrac{u^4}{\gamma^2}
\dfrac{\left(1+2 \kappa_{1}{L_{S}\over r}+{\rm O}(L_{P}^{2})\right)}
{\left(1+2 \kappa_{1}{L_{J}\over \rho}
+{\rm O}(L_{P}^{2})\right)} \sqrt{\dfrac{N(u,\theta)}{D(u,\theta)}}.
\label{(3.22)}
\end{equation}
where (cf. Eqs. (\ref{(2.23)}) and (\ref{(2.24)}))
\begin{equation}
N(u,\theta)=1+\gamma^2 -2\sqrt{\gamma}\left(1+u\cos(\theta)\right)
+4\kappa_1 L_S (r^4 f_2)
+{\rm O}(G^2)=r^4 P_2(u,\theta),
\label{N(u,theta)}
\end{equation}
\begin{equation}
D(u,\theta)=1+2u^2 \cos(\theta)+u^4+4 \kappa_1 (\rho^4 f_1)
+{\rm O}(G^2) = \rho^4 P_1(u,\theta).
\label{D(u,theta)}
\end{equation}
In order to consider the effective gravity version of the
sphere of influence of the planet, we need to
evaluate the small-$u$ behaviour of Eq. (\ref{(3.22)}).
We begin with Eq. (\ref{N(u,theta)}). The last
term of this equation represents a pure effective gravity effect and its expansion
reads as (see Eqs. (\ref{def_f2})-(\ref{(3.20)}))
\begin{equation}
4 \kappa_1 L_S r^4 f_2= 4 \kappa_1 \left(\dfrac{L_S}{\rho}\right)
\left[u \gamma^2 B(\gamma)\right]
=4 \kappa_1 \left(\dfrac{L_S}{\rho}\right)
\Bigl[\left(1+5 \cos^2 (\theta)\right)u^3 + {\rm O} (u^4) \Bigr].
\end{equation}
Therefore, the expansion of Eq. (\ref{N(u,theta)}) leads to (cf. Eq. (\ref{(2.26)}))
\begin{eqnarray}
N(u,\theta)&=& u^2\left(1+3 \cos^{2} (\theta)\right)
+4u^{3}\cos(\theta)+u^3\left( 4 \kappa_1
\dfrac{L_S}{\rho}\right) \left(1+5 \cos^2 (\theta)\right)
\nonumber \\
&+& {\rm O} (u^4) +{\rm O}(G^2).
\end{eqnarray}
In order to obtain the expansion of Eq. (\ref{D(u,theta)}), we first need to
take into account that in our model we can safely write
\begin{equation}
\dfrac{L_S}{r}, \dfrac{L_S}{\rho}, \dfrac{L_J}{\rho}  \ll 1.
\label{ratios}
\end{equation}
This leads to (see Eq. (\ref{(3.17)}))
\begin{eqnarray}
D(u,\theta)&=& 1+ 4 \kappa_1 \left(\dfrac{L_J}{\rho}\right)
+2u^2 \cos(\theta) +4 \kappa_1 u^2
\left(\dfrac{L_J}{\rho}\right) \cos(\theta)
\nonumber \\
&+& 4 \kappa_1 u^3  \left(\dfrac{L_S}{\rho}\right)\cos(\theta)
+ {\rm O} (u^4) +{\rm O}(G^2).
\end{eqnarray}
As a result of the last calculations, we can now evaluate the small-$u$ behaviour of
Eq. (\ref{(3.22)}), that is ruled by the following computation:
\begin{eqnarray}
\; & \; &
\dfrac{u^4}{\gamma^2}  \dfrac{\sqrt{N(u,\theta)}}{\sqrt{D(u,\theta)}}
\nonumber \\
&=& u^4  \dfrac{\sqrt{N(u,\theta)}}
{\left(1+4u \cos(\theta)+{\rm O}(u^2) \right) \left( \sqrt{1+\dfrac{4\kappa_1 L_J}{\rho}}
+{\rm O}(u^2)+{\rm O}(G^2)\right)}
\nonumber \\
& = & \dfrac{u^4}{\sqrt{1+\dfrac{4\kappa_1 L_J}{\rho}}}
\left(1-4u \cos(\theta)+{\rm O}(u^2)
+{\rm O}(G^2) \right)\sqrt{N(u,\theta)}
\nonumber \\
& = & \dfrac{u^5}{\sqrt{1+\dfrac{4\kappa_1 L_J}{\rho}}} \left(1-4u \cos(\theta)+{\rm O}(u^2)
+{\rm O}(G^2) \right)\sqrt{1+3 \cos^2 (\theta)} \nonumber \\
& \times & \Biggl[1+\dfrac{2u}{(1+3\cos^2(\theta))}\left(\cos(\theta) + k_1\dfrac{L_S}{\rho}
+ 5 \kappa_1 \dfrac{L_S}{\rho} \cos^2(\theta) \right)
\nonumber \\
&+& {\rm O}(u^2)+{\rm O}(G^2) \Biggl]
\nonumber \\
& = & \dfrac{u^5 \sqrt{1+3 \cos^2 (\theta)}}{\sqrt{1+4 \kappa_1 (L_J/\rho)}}
\nonumber \\
& \times & \Biggl[1+\dfrac{-2 u
\cos(\theta) \left(1+6 \cos^2(\theta)-5\kappa_1 (L_S/\rho) \cos(\theta)\right)
+2u \kappa_1 (L_S/\rho)}{(1+3\cos^2(\theta))}
\nonumber \\
& + & {\rm O} (u^2) +{\rm O}(G^2)\Biggl].
\end{eqnarray}
Therefore, up to ${\rm O} (u^2)$ and ${\rm O}(G^2)$ terms in the square bracket,
we have (cf. Eq. (\ref{(2.27)}))
\begin{eqnarray}
m^{\prime \,2}&=& u^5  g(\kappa_1) \sqrt{1+3 \cos^2(\theta)}
\nonumber \\
& \times & \Biggl[ 1+\dfrac{-2 u \cos(\theta)
\left(1+6 \cos^2(\theta)-5\kappa_1 (L_S/\rho) \cos(\theta)\right)}
{(1+3\cos^2(\theta))}
\nonumber \\
&+& {2u \kappa_{1}(L_{S}/\rho)\over (1+3 \cos^{2}(\theta))}\Biggl],
\label{m^prime_1}
\end{eqnarray}
where we have defined, up to ${\rm O} (L_P^2)$,
\begin{equation}
g(\kappa_1) \equiv \dfrac{1+2\kappa_1 (L_S/r)}{\left(1+2\kappa_1
(L_J/\rho)\right) \sqrt{1+4 \kappa_1 (L_J/\rho)}}.
\end{equation}
At this stage, we can invert Eq. (\ref{m^prime_1}), finding
\begin{eqnarray}
u^{5} &=& \dfrac{m^{\prime\,2}}{g(\kappa_1) \sqrt{1+3 \cos^2(\theta)}}
\nonumber \\
& \times & \Biggl[ 1+\dfrac{2 u
\cos(\theta) \left(1+6 \cos^2(\theta)-5\kappa_1 (L_S/\rho) \cos(\theta)\right)}
{(1+3\cos^2(\theta))}
\nonumber \\
&-& {2u \kappa_{1}(L_{S}/\rho) \over (1+3 \cos^{2}(\theta))}\Biggl],
\label{u^5}
\end{eqnarray}
whose lowest order solution reads as
\begin{equation}
u= \left( \dfrac{m^{\prime \, 2}}{g(\kappa_1) \sqrt{1+3\cos^{2}(\theta)}}\right)^{1/5}
\equiv u_0^{1/5}.
\end{equation}
Therefore, by solving approximately Eq. (\ref{u^5}) we obtain
\begin{eqnarray}
u & \approx  & u_0^{1/5}+ \dfrac{2}{5} \cos(\theta) u_0^{2/5} \left[\dfrac{1+6 \cos^2
(\theta)-5\kappa_1 (L_S/\rho) \cos(\theta)}{(1+3 \cos^2(\theta))} \right]
\nonumber \\
& - & \dfrac{2}{5} u_0^{2/5} \dfrac{\kappa_1 (L_S/\rho)}{(1+3 \cos^2(\theta))}.
\label{u_approx}
\end{eqnarray}
Since Eq. (\ref{ratios}) implies that $g(\kappa_1) \approx 1$, we obtain
a result similar to the Newtonian case
\begin{equation}
u_0^{1/5} \approx \left( \dfrac{m^{\prime \, 2}}{ \sqrt{1+3\cos^{2}
(\theta)}}\right)^{1/5} < m^{\prime \, 2/5},
\end{equation}
and hence from (\ref{u_approx}) we conclude that the (approximate) equation defining the sphere
of influence of the planet within the effective field theories of gravity picture reads as
\begin{equation}
\rho = r^{\prime} \left( \dfrac{m^{\prime \, 2}}{g(\kappa_1)
\sqrt{1+3\cos^{2}(\theta)}}\right)^{1/5}.
\label{effective_sphere_of_influence}
\end{equation}
From the above equation it is clear that the effective field theory framework produces a tiny
variation of the radius of the sphere of influence by means of the factor $1/g(\kappa_1)$.

\subsection{Comet trajectories in effective and Newtonian gravity} \label{Sec_traj_EFT-Newt}

In this section we show the trajectories followed by the comet both in Newtonian and in
effective field theory of gravity. As we will see, the corrections introduced in the
comet motion by the effective picture turn out to be too tiny to be testable.

According to the effective field theory prescriptions (\ref{(3.1)})-(\ref{(3.4)}) the
heliocentric motion is described by the set of differential equations
(hereafter we neglect terms ${\rm O}(L_P^2)$)
\begin{eqnarray}
\; & \; &
\left[\dfrac{d^2}{dt^2} +\dfrac{G m_{\odot}}{r^3} \left(1+2 \kappa_1\dfrac{L_S}{r}\right)
\right]x_k
\nonumber \\
&=& G m^{\prime} \left[ \dfrac{\left(x^\prime_k-x_k \right)}{\rho^3} \left(1+2
\kappa_1\dfrac{L_J}{\rho}\right)-\dfrac{x^\prime_k}{r^{\prime 3}}\left(1+\dfrac{2
\kappa_1L_S}{r^\prime}\right) \right],
\nonumber \\
& \; & (k=1,2,3),
\label{heliocentric_effective}
\end{eqnarray}
whereas the motion of Jupiter is given in terms of Eqs.
(\ref{(2.5)})-(\ref{(2.7)}) provided
that Eq. (\ref{newtonian_Omega}) is subjected to the change
\begin{equation}
\Omega^2 \rightarrow \tilde{\Omega}^2= {G(m_{\odot}+m') \over {r'}^{3}} \left(1
+ 2 \kappa_1 \dfrac{L_S}{r^\prime} \right).
\label{Omega_eff}
\end{equation}
Therefore, Eqs. (\ref{(2.5)})-(\ref{(2.7)}) are replaced by
\begin{equation}
\left(\dfrac{d^2}{dt^2} + \tilde{\Omega}^2\right)x^\prime_k=0, \;\;\;\; (k=1,2,3).
\label{effective_Jupiter}
\end{equation}

\begin{figure}
\centering
\includegraphics[height=6cm]{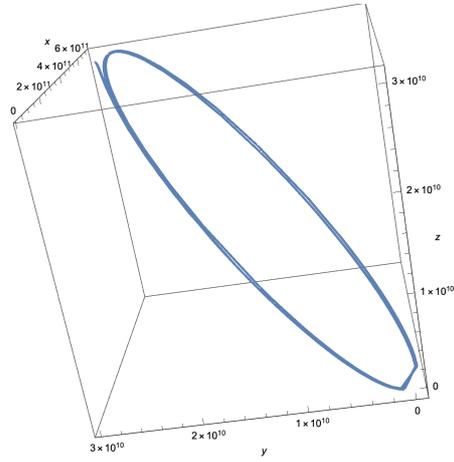}  
\caption{Heliocentric comet motion in Newtonian gravity.}
\label{fig_helioc_3d_a}
\end{figure}

\begin{figure}
\centering
\includegraphics[height=6cm]{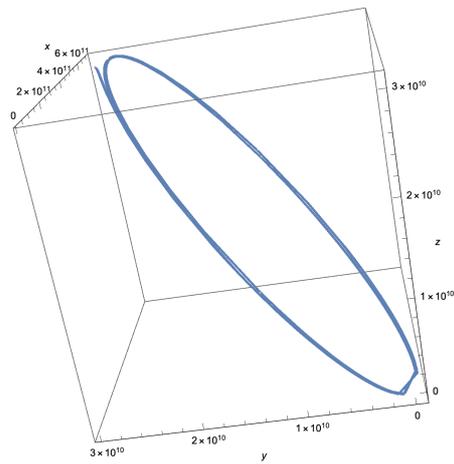}  
\caption{Heliocentric comet motion in effective field theory of gravity.}
\label{fig_helioc_3d_b}
\end{figure}

\begin{figure}
\centering
\includegraphics[height=6cm]{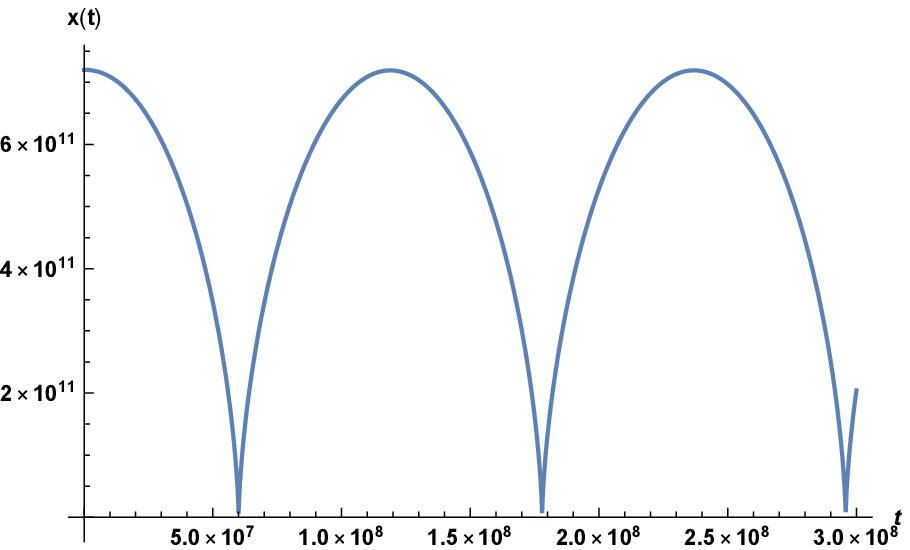} 
\caption{The component $x(t)$ of the solution  $\vec{r}(t)$ of the effective field theory 
equations (\ref{heliocentric_effective})-(\ref{effective_Jupiter}).}
\label{figura2_helioc_a}
\end{figure}

\begin{figure}
\centering
\includegraphics[height=6cm]{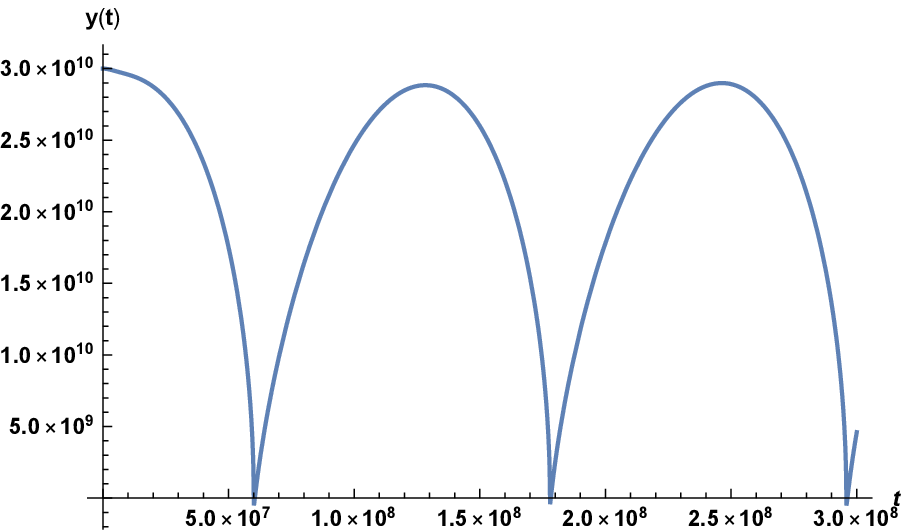} 
\caption{The component $y(t)$ of the solution  $\vec{r}(t)$ of the effective field theory 
equations (\ref{heliocentric_effective})-(\ref{effective_Jupiter}).}
\label{figura2_helioc_b}
\end{figure}

\begin{figure}
\centering
\includegraphics[height=6cm]{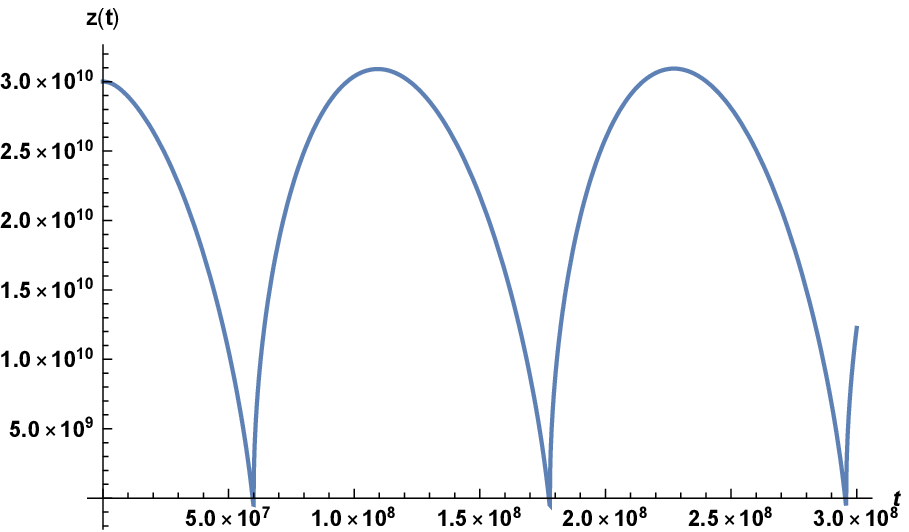} 
\caption{The component $z(t)$ of the solution  $\vec{r}(t)$ of the effective field theory 
equations (\ref{heliocentric_effective})-(\ref{effective_Jupiter}).}
\label{figura2_helioc_c}
\end{figure}

The plots of the heliocentric trajectory of the comet in Newtonian gravity (see Eqs.
(\ref{(2.2)})-(\ref{(2.7)})) and in effective field theory of gravity are 
displayed in Figs. \ref{fig_helioc_3d_a} and \ref{fig_helioc_3d_b}, respectively.  
In obtaining these figures, Eqs. (\ref{(2.2)})-(\ref{(2.7)}) and
their effective gravity counterpart (\ref{heliocentric_effective})-(\ref{effective_Jupiter})
have been integrated by choosing the following initial conditions at the time $t_0=0$:
$\vec{r}(t_0)= (7.2 \times 10^{11} \,{\rm m},3 \times 10^{10} \,{\rm m},3 \times 10^{10}\,
{\rm m})$, $\dfrac{d}{dt}\vec{r}(t_0)= (80\, {\rm m/s},0,0)$, $\vec{r}^{\,\prime}(t_0)
=(7.78 \times 10^{11} \,{\rm m},0,0)$, $\dfrac{d}{dt}\vec{r}^{\,\prime}(t_0)
= (0,1.3 \times 10^4\,{\rm m/s},0)$. The solution
$\vec{r}(t)=(x(t),y(t),z(t))$ of Eqs.
(\ref{heliocentric_effective})-(\ref{effective_Jupiter})
is drawn in Figs. \ref{figura2_helioc_a}-\ref{figura2_helioc_c}.

Bearing in mind Eqs. (\ref{(3.1)}), (\ref{(3.2)}), (\ref{(3.9)}), and (\ref{(3.10)}), within the
effective field theory domain the jovicentric motion of the comet is given by
\begin{eqnarray}
\; & \; &
\left[\dfrac{d^2}{dt^2} +\dfrac{G m^\prime}{\rho^3} \left(1+2 \kappa_1\dfrac{L_J}{\rho}\right)
\right]\xi_k
\nonumber \\
&=& G m_{\odot} \left[\dfrac{x^\prime_k}{r^{\prime 3}} -\dfrac{x_k}{r^3}
+ 2 \kappa_1 \left( \dfrac{L_S}{r^\prime}\dfrac{x^\prime_k}{r^{\prime 3}}
- \dfrac{L_S}{r}\dfrac{x_k}{r^3}\right)\right],
\nonumber \\
& \; & (k=1,2,3)
\label{jovicentric_effective}
\end{eqnarray}
along with Eqs. (\ref{Omega_eff}) and (\ref{effective_Jupiter}).

\begin{figure}
\centering
\includegraphics[height=6cm]{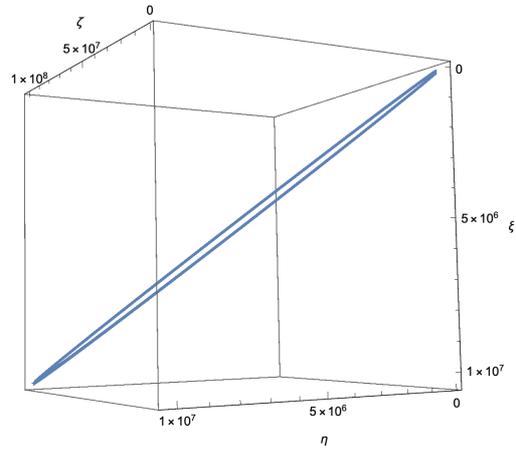}
\caption{Jovicentric comet motion in Newtonian gravity.}
\label{fig_jovic_3d_a}
\end{figure}

\begin{figure}
\centering
\includegraphics[height=6cm]{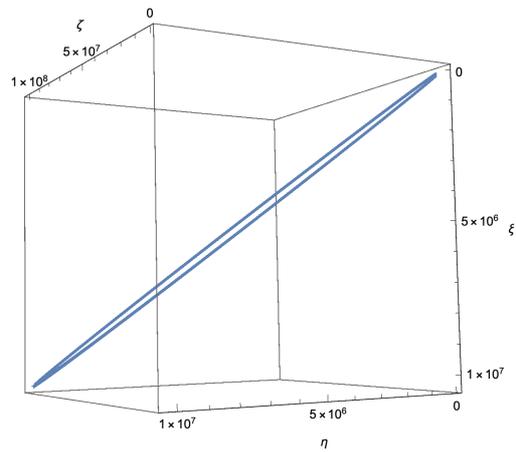}
\caption{Jovicentric comet motion in effective field theory of gravity.}
\label{fig_jovic_3d_b}
\end{figure}

\begin{figure}
\centering
\includegraphics[height=6cm]{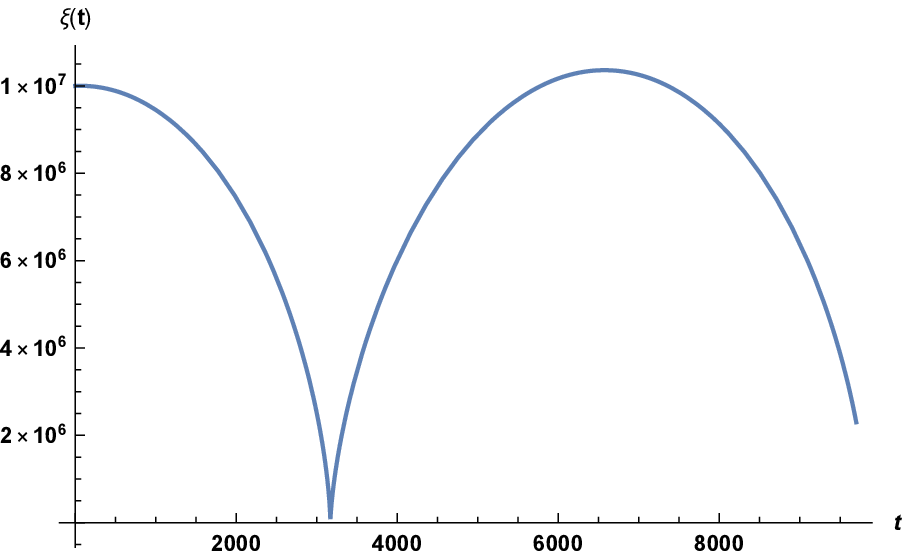}
\caption{The component $\xi(t)$ of the solution $\vec{\rho}(t)$ of the effective field theory
equations (\ref{Omega_eff})-(\ref{jovicentric_effective}).}
\label{fig4_jovic_a}
\end{figure}

\begin{figure}
\centering
\includegraphics[height=6cm]{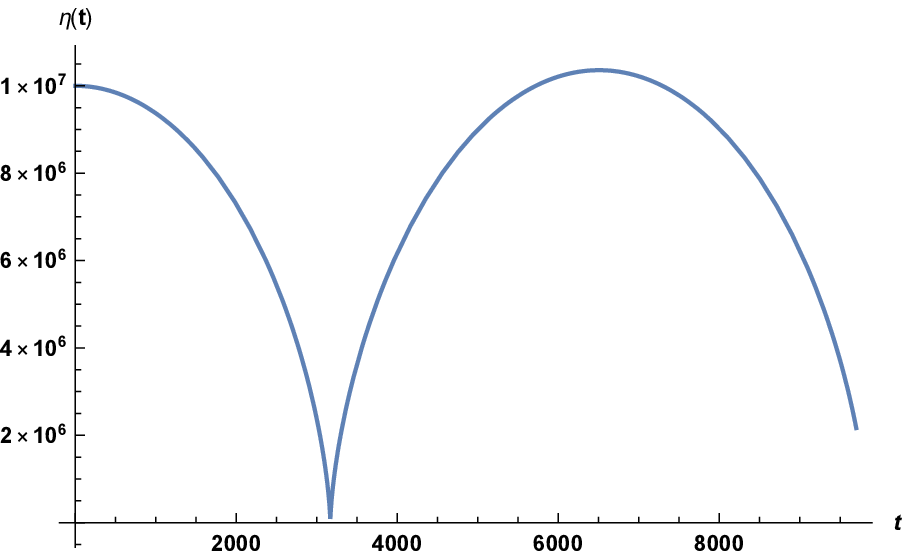}
\caption{The component $\eta(t)$ of the solution $\vec{\rho}(t)$ of the effective field theory
equations (\ref{Omega_eff})-(\ref{jovicentric_effective}).}
\label{fig4_jovic_b}
\end{figure}

\begin{figure}
\centering
\includegraphics[height=6cm]{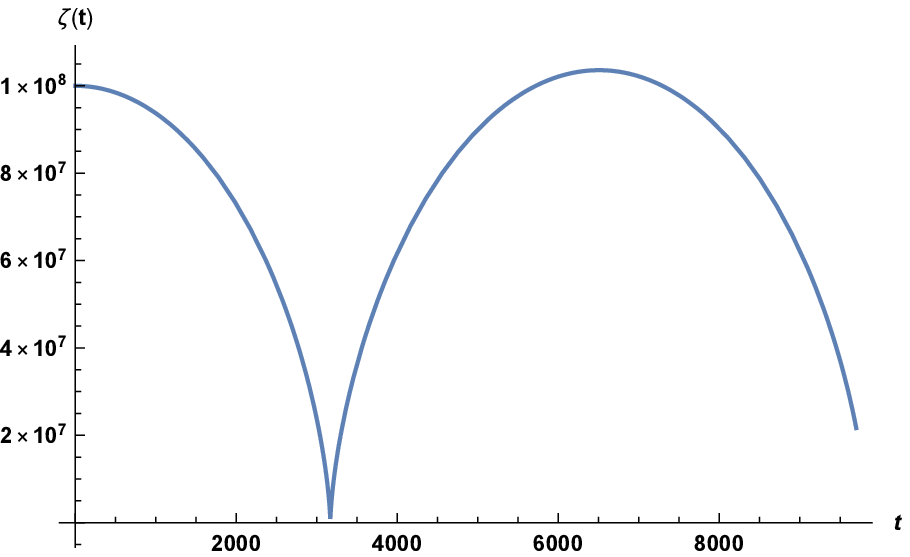}
\caption{The component $\zeta(t)$ of the solution $\vec{\rho}(t)$ of the effective field theory
equations (\ref{Omega_eff})-(\ref{jovicentric_effective}).}
\label{fig4_jovic_c}
\end{figure}

The jovicentric motion of the comet both in Newtonian
(cf. Eqs. (\ref{(2.5)})-(\ref{(2.7)}) and
(\ref{(2.9)})-(\ref{(2.11)})) and effective gravity is depicted in Figs. 
\ref{fig_jovic_3d_a} and \ref{fig_jovic_3d_b}. Equations (\ref{(2.5)})-(\ref{(2.7)}),
(\ref{(2.9)})-(\ref{(2.11)}) and their effective gravity counterpart  (\ref{Omega_eff})
-(\ref{jovicentric_effective}) have been integrated by employing the following initial conditions
at the time $t_0=0$: $\vec{\rho}(t_0)= (1.0 \times 10^{7} \,{\rm m},1.0 \times 10^{7} \,{\rm m},1.0
\times 10^{8}\, {\rm m})$, $\dfrac{d}{dt}\vec{\rho}(t_0)= (80\, {\rm m/s},0,0)$,
$\vec{r}^{\,\prime}(t_0)=(7.78 \times 10^{11} \,{\rm m},0,0)$,
$\dfrac{d}{dt}\vec{r}^{\,\prime}(t_0)= (0,1.3 \times 10^4\,{\rm m/s},0)$
The solution $\vec{\rho}(t)=(\xi(t),\eta(t),\zeta(t))$ can be read 
from Figs. \ref{fig4_jovic_a}-\ref{fig4_jovic_c}.

In obtaining Figs. \ref{fig_helioc_3d_a}-\ref{fig4_jovic_c} the initial conditions regarding the
motion of Jupiter reflect its average orbital speed and its average distance from the Sun. With
these choices, the radius $\rho_0$ evaluated at the initial time $t_0=0$ of the (classical)
sphere of influence of Jupiter reads as (restoring the solar mass $m_{\odot}$)
\begin{equation}
\rho_0 = \left(\dfrac{m^\prime}{m_{\odot}}\right)^{2/5} \, r^\prime (t_0)
= 4.82 \times 10^{10}\, {\rm m}.
\end{equation}
We have analysed several alternatives for the initial conditions of the comet motion.
In all cases, the differences between the classical and the effective
regime are unperceivable.
The initial conditions featuring the heliocentric motion sketched
in Figs. \ref{fig_helioc_3d_a}-\ref{figura2_helioc_c} are such that the initial distances
comet-Sun $r_{\rm h}(t_0)$ and comet-Jupiter $\rho_{\rm h}(t_0)$ are
\begin{equation}
r_{\rm h}(t_0) = 7.21 \times 10^{11}\, {\rm m}, \;
\rho_{\rm h}(t_0) = 7.19 \times 10^{10}\, {\rm m},
\end{equation}
respectively, with the parameter $u_{\rm h}(t_0)$ given by (cf. Eq. ({\ref{(2.20)}}))
\begin{equation}
u_{\rm h}(t_0)=\dfrac{\rho_{\rm h}(t_0)}{r^\prime(t_0)}=0.092.
\end{equation}

On the other hand, the jovicentric motion (Figs. \ref{fig_jovic_3d_a}-\ref{fig4_jovic_c}) 
is characterized by (the reader should be aware that,
in Figs. \ref{fig_jovic_3d_a} and \ref{fig_jovic_3d_b}, 
the closed orbit results from a particular choice of initial
conditions, while for other choices an open orbit is instead obtained)
\begin{equation}
r_{\rm j}(t_0) = 7.78 \times 10^{11}\, {\rm m}, \;
\rho_{\rm j}(t_0) = 1.01 \times 10^{8}\, {\rm m},
\end{equation}
so that
\begin{equation}
u_{\rm j}(t_0)=\dfrac{\rho_{\rm j}(t_0)}{r^\prime(t_0)}=1.30 \times 10^{-4}.
\end{equation}

The analysis performed in this section clearly shows that the corrections to the comet motion
resulting from the effective field theory approach are negligible. This agrees with
the outcome of the previous section, where Eq. (\ref{effective_sphere_of_influence})
describing the sphere of influence of the planet indicates a tiny departure from the classical case.

\section{Concluding remarks} \label{Sec_conclusions}

For the first time in the literature, the dynamics of a comet has been investigated within
a framework where the dimensionless weight factors for gravitational radii are
obtained from quantum field theory.
We have proposed that the effective-gravity regime should rely on the set of original prescriptions
(\ref{(3.1)})-(\ref{(3.4)}) and we have found that our
model predicts very tiny departures from the
Newtonian picture. This is witnessed both by Eq. (\ref{effective_sphere_of_influence})
defining the quantum corrected sphere of influence of Jupiter and by the comet
trajectories of Figs. \ref{fig_helioc_3d_a}, \ref{fig_helioc_3d_b}, 
\ref{fig_jovic_3d_a} and \ref{fig_jovic_3d_b}.

The next task is of course the analysis of cometary motions in general relativity.
Within this framework, a full classification of the orbits can be found in Ref.
\cite{F1}, devoted to the orbital dynamics in the post-Newtonian planar
circular restricted Sun-Jupiter system. Moreover, one should exploit the Hamiltonian
for the three-body problem at second post-Newtonian order in the
Arnowitt-Deser-Misner gauge obtained in Ref. \cite{F2}, and also the important
work on the three-body equations of motion in Ref. \cite{F3}, and the results on
post-Newtonian dynamics for time-dependent metrics \cite{F4}.

Third, one should bear in mind
that three-body systems display chaotic behaviour over
sufficiently long times \cite{Wanex}. This implies that there might exist critical
combinations of some parameters, and in the neighborhood of such critical values, even a
very small perturbation could give rise to orbits that differ a lot from each other.
From this point of view, even the detailed calculations of Sec. \ref{Sec_perturb_EFT} are not just of
academical interest. Furthermore, a numerical or analytic investigation of non-gravitational
effects such as evaporation of the comet's head, and radiation pressure, might prove useful.

We are aware of the fact that the description of cometary dynamics 
does not  need in principle a quantum description of the low-energy 
regime of the gravitational field. In spite of that, we strongly 
believe that this represents one of the reasons for which our paper 
can be of some interest. Indeed, we have shown that quantum gravity 
leads to some low-energy modifications even in  a purely classical 
and observationally well established setting like cometary dynamics. 
This circumstance can give rise to a valuable resource since we can 
devise a favourable scenario where Eqs. (3.32) and  
(\ref{effective_sphere_of_influence}) 
can be tested and hence used in order to constrain the factors 
$\kappa_1$ and $\kappa_2$ occurring in Eq. (\ref{(1.1)})\footnote{We 
recall that in effective field theory of gravity the corrections 
$\kappa_1$ and $\kappa_2$ occurring in Eq. (\ref{(1.1)}) are intimately 
intertwined and the output of one factor affects the value of the other.}.  
This is not  a pointless issue since in the literature there is not a 
general agreement on the value assumed by the term $\kappa_2$ occurring 
in Eq. (\ref{(1.1)}) (see \emph{e.g.} Tab. I in Ref. \cite{Bargueno2017} 
where our $\kappa_2$ is indicated  with the Greek letter $\gamma$). 
This means that the research of observables of low-energy quantum gravity 
constitutes a context where further efforts must be made. It is in this 
spirit that the content of our paper, representing the natural continuation 
of a research project undertaken by two of us some years ago, should be considered. 

\section*{Acknowledgments}
G. E. is grateful to Dipartimento di Fisica ``Ettore Pancini'' for hospitality and support,
and to G. Longo for correspondence. E. B. is grateful to V. De Falco for useful discussions.

\end{document}